\documentclass{amsart}
\usepackage{mathrsfs,amsmath,amssymb}
\def\alg#1{{\mathcal #1}}\def\set#1{{\sf #1}}\def\sH{\set{H}}\def\Tr{\operatorname{Tr}}
\def\d{\operatorname{d}}
\def\<{\langle}\def\>{\rangle}\def\geq{\geqslant}\def\leq{\leqslant}\def\St#1{\alg{S(#1)}}
\def\Pinv#1#2{\alg{P_{#2}(\set{#1})}}
\newtheorem{lemma}{Lemma}
\newtheorem{corollary}{Corollary}
\def\Proof{\medskip\par\noindent{\bf Proof. }}\def\qed{$\,\blacksquare$\par} 
\def\ea#1{{\tt #1}}\def\hpg#1{{\tt #1}}
\begin{document}
\title{Optimal estimation of quantum observables}
\author[Giacomo Mauro D'Ariano, Vittorio Giovannetti, and Paolo Perinotti]{Giacomo Mauro
  D'Ariano${}^{1,a}$, Vittorio Giovannetti${}^{2,b}$, Paolo Perinotti${}^{1,c}$}
\address{${}^1$QUIT Group, Dipartimento di Fisica ``A. Volta'', via Bassi 6, 27100 Pavia,
  Italy ${}^2$Scuola Normale Superiore, Piazza dei Cavalieri 7, 56126 Pisa, Italy\vskip 1em
${}^a$\ea{dariano@unipv.it},
${}^b$\ea{v.giovannetti@sns.it},
${}^c$\ea{perinotti@fisicavolta.unipv.it},\vskip 1em
${}^\dag$\hpg{http://www.qubit.it}
}
\date{\today}


\begin{abstract}
  We consider the problem of estimating the ensemble average of an observable on an ensemble of
  equally prepared identical quantum systems. We show that, among all kinds of measurements
  performed jointly on the copies, the optimal unbiased estimation is achieved by the usual
  procedure that consists in performing independent measurements of the observable on each system
  and averaging the measurement outcomes.\\


\end{abstract}
\maketitle
\section{Introduction}
\label{sec:intro}
The astonishing precision of measurements currently available in quantum optics \cite{Meystre83}
along with the growing demand of quantum devices of the new information technology
\cite{pop,Nielsen2000} have revived the interest in the theory of quantum measurements \cite{Busch}.
The outcome statistics of a quantum measurement for all possible input states is described by a
positive operator valued measure (POVM).  The general optimization approach of Quantum Estimation
Theory \cite{helstrom} is to maximize over all possible POVM's an appropriate cost function, which
depends on the context and on the specific use of the measurement. The output statistics can then be
improved by using multiple copies of the same quantum system, all prepared in the same state, and
performing a suitable {\em ensemble measurement} over the copies.

The experimental complexity of ensemble measurements is roughly classified by dividing them into
three main categories: a) {\em independent}, b) {\em separable}, and c) {\em entangled}
measurements.  Category a) is described by tensor products of independent POVM's; b) by POVM's with
separable elements only; c) by POVM's where some elements are entangled. Notice that the
separability of POVM's generally does not correspond to a physical separability of measuring
apparatuses\footnote{There exist separable measurements that cannot be performed by separate
  measuring apparatuses, i.~.e. by local operations and classical communication (LOCC)}, and this
classification remains essentially mathematical in nature. However, at least one can say that
category b) contains all {\em adaptive} measurements (in which the choice of the measuring apparatus
on the $n$th copy depends on the outcomes of previous measurements), whereas category c) contains
those measurements that need quantum interactions between copies, implying that all copies during
the measuring time must be at the same physical location, or, otherwise, that a ``quantum memory''
is available.

Among the three categories of ensemble measurements, the category c) of entangled POVM's discloses
the full exponential growth of the Hilbert space dimension versus the number of copies $N$ for a
virtually unlimited optimization of the statistical efficiency of the measurement, with the
possibility of largely surpassing the performance of categories a) and b)
\cite{buggec,Chefles,lopar,olevoasymptotic}.  Indeed, over the last few years, it has been
recognized that entangled measurements are usually more efficient than independent measurements, and
the optimal measurement scheme is almost always entangled
\cite{refframe,phase,covmeas,infolocvsglob}.  However, in some situations it has been also shown
that asymptotically for $N\to\infty$ an equivalently optimal estimation may be achieved using just
independent measurements over the copies\cite{Ballester,Ballester2,Ballester3,Ballester4,Vittorio}.

In the above scenario it is natural to ask if the {\em canonical procedure} of averaging the
outcomes of repeated measurements of an observable $A$ over equally prepared systems is the best way
of estimating the ensemble average $\<A\>$ of $A$, or, instead, if a joint entangled measurement
over the copies can improve the estimation. As we will see it turns out that the canonical procedure
is indeed optimal, however, the derivation of this result is non trivial, and opens a general
warning against easy assumptions and generalizations when evaluating statistical efficiencies of
ensemble measurements.

Let's be more precise, and fix precisely the scenario of the quantum estimation. Suppose one has a
finite number $N$ of equally prepared distinguishable identical $d$-dimensional quantum systems,
which are described by the state $\rho^{\otimes N}$, and one wants to estimate the ensemble average
$\<A\>_\rho\equiv\Tr[\rho A]$ of the observable $A$. Suppose now that one has unlimited technology
at disposal, including measuring apparatus that can achieve any desired entangled POVM on all $N$
systems jointly.  The question is: which is the best measuring apparatus to choose in order to
estimate $\<A\>_\rho$ with the minimum statistical error?  What we will prove in the present paper
is that the best estimation strategy is just the canonical procedure, which consists in averaging the
outcomes of repeated measurements of the observable $A$ over the equally prepared quantum systems.
\section{Permutationally invariant polarization identities}

\par In the derivation of our main result the following lemma will play a crucial role.

\begin{lemma}\label{permi}
Any permutationally invariant operator $X$ on $\sH^{\otimes N}$ is completely determined by all
ensemble averages  $\Tr[X\rho^{\otimes N}]$ on identical equally prepared systems.
\end{lemma}

\Proof The statement of the lemma is equivalent to the following logical implication
\begin{equation}\label{eq:statement}
X\in\Pinv{H}{N},\;\;\forall\rho\in\St{H},\quad \Tr[X\rho^{\otimes N}]=0\quad\Rightarrow\quad X=0,
\end{equation}
where $\St{H}$ denotes the set of states on $\sH$, and $\Pinv{H}{N}$ the algebra of permutationally
invariant operators on $\sH^{\otimes N}$. Indeed, statement (\ref{eq:statement}) is equivalent to
the statement that if $\Tr[X\rho^{\otimes N}]=\Tr[Y\rho^{\otimes N}]$ for all states $\rho$, then $X\equiv Y$. 

Consider the following special states of the form
\begin{equation}
\rho_\lambda=\sum_{j=1}^N\lambda_j|\psi_j\>\<\psi_j|,\qquad\lambda_j>0,\;\lambda_j\neq \lambda_i,\, i\neq j,
\end{equation}
with $\{\psi_j\}_N$ any set of $N$ unequal states (not necessarily orthogonal). The trace
$\Tr[X\rho_\lambda^{\otimes N}]$ is a polynomial in $\prod_{j=1}^{N}\lambda_j^{x_j}$, with
$\sum_{j=1}^Nx_j=N$ and $x_j\geq0$ integers. Now for $\Tr[X\rho_\lambda^{\otimes N}]=0$ all
coefficients of the polynomial must vanish.  In particular, the coefficient of
$\prod_{j=1}^N\lambda_j$ is given by
\begin{equation}
\sum_\sigma \<\psi_1|\dots\<\psi_N|\Pi_\sigma X\Pi_\sigma^\dag|\psi_1\>\dots|\psi_N\>\equiv0,
\label{factpolariz}
\end{equation}
where $\Pi_\sigma$ are the permutations of the $N$ systems. By hypothesis we have $\Pi_\sigma
X\Pi_\sigma^\dag=X$, then the vanishing of $\Tr[X\rho_\lambda^{\otimes N}]$ for all states $\rho_\lambda$ implies
\begin{equation}
\<\psi_1|\dots\<\psi_N|X|\psi_1\>\dots|\psi_N\>=0,
\end{equation}
for all sets $\{\psi_j\}_N$. If we take $|\psi_k\>=\alpha |\phi\>+\beta|\phi_\perp\>$, by
arbitraryness of $\alpha$ and $\beta$ we have 
\begin{equation}
\begin{split}
&\<\psi_1|\dots\<\phi|\dots\<\psi_N|X|\psi_1\>\dots|\phi\>\dots|\psi_N\>=\\
&\<\psi_1|\dots\<\phi|\dots\<\psi_N|X|\psi_1\>\dots|\phi_\perp\>\dots|\psi_N\>=\\
&\<\psi_1|\dots\<\phi_\perp|\dots\<\psi_N|X|\psi_1\>\dots|\phi\>\dots|\psi_N\>=\\
&\<\psi_1|\dots\<\phi_\perp|\dots\<\psi_N|X|\psi_1\>\dots|\phi_\perp\>\dots|\psi_N\>=0\,.
\end{split}
\end{equation}
By repeating the same argument for different values of $k$ and
choosing $\phi$ and $\phi_\perp$ as all possible elements of an
orthonormal basis $\{\phi_j\}$ we get
\begin{equation}
\<\phi_{j_1}|\dots\<\phi_{j_N}|X|\phi_{k_1}\>\dots|\phi_{k_N}\>=0\,,\quad\forall\{j_i\}\,,\{k_i\}\,.
\end{equation}
Since all the matrix elements on an orthonormal basis of $X$ are null, one has that $X\equiv0$.
\qed

Notice that the proof of the previous lemma contains the following interesting corollary
\begin{corollary}
For any permutation invariant operator $X$ on $\sH^{\otimes N}$ the diagonal elements on factorized states
completely determine $X$
\end{corollary}
This is a kind of {\em factorized} polarization identity for permutation invariant operators.\par

\section{The main result}
Let's now come back to the original problem of determining the optimal measurement for estimating the
ensemble average of an observable. Consider a generic joint POVM $P(r)$ on $\sH^{\otimes N}$, with
outcome $r$ providing an estimate of the expectation $\<A\>_\rho$ of the observable $A$ on $N$
identical systems all in the same state $\rho$. Clearly, one has $\lambda_m\leq r\leq\lambda_M$,
with $\lambda_m$ and $\lambda_M$ minimum and maximum eigenvalues of $A$, respectively. The POVM 
 $P(r)$ provides an estimate of the expectation $\<A\>_\rho$ if the conditional probability
 $p(r|\rho)$ of estimating expectation value $r$ for actual value $\Tr[A\rho]$ is expressed via the
Born rule as follows
\begin{equation}\label{conditional}
p(r|\rho)\d r=\Tr[ P(r)\rho^{\otimes N}]\d r.
\end{equation}
Since the state $\rho^{\otimes N}$ is permutation invariant, we can consider permutation invariant
POVM's. Indeed, using invariance of  $\rho^{\otimes N}$ under permutations, one has
\begin{equation}
\begin{split}
p(r|\rho)\d r&=\frac1{N!}\sum_\sigma\Tr[\Pi_\sigma\rho^{\otimes N}\Pi_\sigma^\dag  P(r)]\d r=\\
&\Tr\left[\rho^{\otimes N}\frac1{N!}\sum_\sigma\left(\Pi_\sigma^\dag  P(r)\Pi_\sigma\right)\right]\d r=\Tr[\Pi'(r)\rho^{\otimes N}]\,,
\end{split}
\end{equation}
where the POVM
\begin{equation}
P'(r)\equiv\frac1{N!}\sum_\sigma\Pi_\sigma^\dag P(r)\Pi_\sigma
\end{equation}
is permutation invariant by construction. This means that for any POVM there is a permutation
invariant one giving the same probability distributions for all states $\rho^{\otimes N}$.
Therefore, without loss of generality, in the following we can assume that $ P(r)$ is permutation
invariant. We will consider now the case in which the POVM is {\em unbiased}, namely that the
averaging over $r$ coincides with the value to be estimated. Mathematically this means that for all
states $\rho$ the following identity holds
\begin{equation}
\int_{\lambda_m}^{\lambda_M}\d r\, rp(r|\rho)=\Tr[A\rho]\,.
\end{equation}
The statistical error in the estimate is given by the r.m.s. of the
probability distribution
\begin{equation}
\epsilon_N(A)\doteq\left[\int_{\lambda_m}^{\lambda_M}\d r\,
  (r-\<A\>_\rho)^2p(r|\rho)\right]^\frac12\,, 
\end{equation}
which for unbiased estimation equals
\begin{equation}\label{error}
\epsilon_N(A)\doteq\left[\int_{\lambda_m}^{\lambda_M}\d r\,\left[r^2
    p(r|\rho)\right]-\<A\>_\rho^2\right]^\frac12\,. 
\end{equation}
Since the only part which depends on the POVM is the conditional probability $ p(r|\rho)$, the
optimization of the error resorts to minimize the quantity
\begin{equation}
\int_{\lambda_m}^{\lambda_M}\d r\,r^2 \Tr[ P(r)\rho^{\otimes N}]\,,\label{mini}
\end{equation}
with the constraints
\begin{align}
&\int_{\lambda_m}^{\lambda_M}\d r\, P(r)=I,\label{norm}\\
&\int_{\lambda_m}^{\lambda_M}\d r\,r\Tr[ P(r)\rho^{\otimes N}]=\<A\>_\rho.\label{unbi}
\end{align}
Using the following identity
\begin{equation}\label{Theta1}
\<A\>_\rho=\Tr\left[\rho^{\otimes N}\frac1{N!}\sum_\sigma \Pi_\sigma(A\otimes I^{\otimes (N-1)})\Pi_\sigma^\dag\right]=\frac1N\sum_{k=1}^NA^{(k)}
\end{equation}
with $A^{(k)}\doteq I^{\otimes(k-1)}\otimes A\otimes I^{\otimes N-k}$,
by virtue of Lemma \ref{permi} we can recast Eq.~\eqref{unbi} as follows
\begin{equation}\label{Theta2}
\int_{\lambda_m}^{\lambda_M}\d r\,r P(r)=\frac1N\sum_{k=1}^NA^{(k)}\doteq\Theta.
\end{equation}
\par The operator $\Delta\geq0$ defined as
\begin{equation}
\Delta\doteq\int_{\lambda_m}^{\lambda_M}\d r\,r^2 P(r)\,,
\end{equation}
allows to re-express the statistical error as follows
\begin{equation}
{\epsilon_N(A)}^2=\Tr[\Delta\rho^{\otimes N}]-\<A\>_\rho^2\,.
\end{equation}
In the representation in which $\Delta$ is diagonal, the constraints (\ref{norm}) and
(\ref{unbi}) become
\begin{align}
&\int_{\lambda_m}^{\lambda_M}\d r\,P(r)_{lk}=\delta_{lk}\label{einorm}\\
&\int_{\lambda_m}^{\lambda_M}\d r\,rP(r)_{lk}=\Theta_{lk}\label{eiav}
\end{align}
whereas the error (\ref{error}) becomes
\begin{equation}
{\epsilon_N(A)}^2=\sum_n(\rho^{\otimes N})_{nn}\int_{\lambda_m}^{\lambda_M}\d r\,r^2P(r)_{nn}-\<A\>_\rho^2.\label{eierr}
\end{equation}
From Eqs. \eqref{einorm} and \eqref{eiav} it follows that the diagonal elements $P(r)_{nn}$ are
probability densities versus $r$ over $[\lambda_m,\lambda_M]$, with average $\Theta_{nn}$, and
upon denoting their variance by $\sigma^2_n$, we can write
\begin{equation}
{\epsilon_N(A)}^2=\sum_n(\rho^{\otimes N})_{nn}(\sigma_n^2+\Theta_{nn}^2)-\<A\>_\rho^2.
\end{equation}
Therefore, ${\epsilon_N(A)}^2$ is minimized by taking $\sigma^2_n=0$, corresponding to 
$P(r)_{nn}\equiv\delta(\Theta_{nn}-r)$. This implies that the outcomes of the optimal POVM are
actually discrete, corresponding to $r_n=\Theta_{nn}$. In this discrete version, the POVM has
$P(r_n)_{nm}=\delta_{nm}$ [which also implies that $\Theta_{nm}=\delta_{nm}\Theta_{nn}$ via
Eq. (\ref{eiav})], namely $P(r_n)$ is projection-valued on the $n$th eigenvector of $\Delta$, 
[when it happens that $\Theta_{nn}=\Theta_{mm}$ for some $m\neq n$, then the projector has rank
equal to the number of equal diagonal elements]. We have finally
\begin{equation}
{\epsilon_N(A)}^2=\sum_n(\rho^{\otimes N})_{nn}\Theta_{nn}^2-\<A\>_\rho^2\,.
\end{equation}
Moreover, we have
\begin{align}
&I=\sum_nP(r_n)\\
&\Theta=\sum_n\Theta_{nn}P(r_n)\\
&\Delta=\sum_n\Theta_{nn}^2P(r_n)\,.
\end{align}
Since optimization makes $\Theta$ and $\Delta$ jointly diagonal, one has
$\sum_n\Theta_{nn}^2(\rho^{\otimes N})_{nn}=\Tr[\Theta^2\rho^{\otimes N}]$, and using
Eqs. (\ref{Theta1}) and (\ref{Theta2}) we can write the following expression for the minimal error
\begin{equation}
{\epsilon_N(A)}^2=\frac1{N^2}\sum_{i,j=1}^N\Tr[A^{(i)}A^{(j)}\rho^{\otimes N}]-\<A\>_\rho^2\,.
\end{equation}
Notice that the sum in the first term contains $N$ terms with $i=j$ equal to $\Tr[A^2\rho]$ and 
and $N(N-1)$ with $i\neq j$ equal to $\<A\>_\rho^2$, resulting in
\begin{equation}
\epsilon_N(A)=\sqrt{\frac{\<A^2\>_\rho-\<A\>_\rho^2}{N}},
\end{equation}
namely the optimal error equals the statistical error occurring when measuring $A$ separately on all
the identical quantum systems in the state $\rho$, and then averaging. Indeed, the optimal POVM
coincides with the spectral resolution of $\Theta=\frac{1}{N}\sum_nA^{(n)}$ on $\sH^{\otimes N}$.

\section*{Acknowledgements}
Giacomo Mauro D'Ariano and PP acknowledge financial support by the EC under the program ATESIT
(Contract No. IST-2000-29681), and by MIUR under programs Cofinanziamento 2003 and FIRB 2001. VG
acknowledges financial support by MIUR through Cofinanziamento 2003.


\begin{thebibliography}{99}
\bibitem{Meystre83} {\em Quantum Optics, Experimental Gravity, and Measurement Theory},
  ed. P. Meystre and M. O. Scully, (Plenum Press, New York and London 1983)
\bibitem{pop} {\em Introduction to Quantum Computation and 
Information}, ed. by H.-K. Lo, S. Popescu, and T. Spiller (World 
Scientific, Singapore, 1998).
\bibitem{Nielsen2000} I. L. Chuang and M. A. Nielsen, {\em Quantum Information and Quantum
    Computation} (Cambridge University Press, Cambridge, 2000).
\bibitem{Busch} P. Busch, P. J. Lahti, and P. Mittelstaedt, {\em The
Quantum Theory of Measurement}, Lecture Notes in Physics Vol. 2
\bibitem{helstrom}C. W. Helstrom, {\it Quantum detection and estimation theory} (Academic Press, New
  York, 1976). 
\bibitem{buggec} R. Derka, V. Bu\u zek, and A. K. Ekert, Phys. Rev.
  Lett. {\bf 80}, 1571 (1998).
\bibitem{Chefles} A. Chefles, Phys. Rev. A {\bf 64}, 062305 (2001).
\bibitem{lopar} G. M. D'Ariano, and P. Lo Presti, M. G. A. Paris, Phys. Rev. Lett. {\bf 87}, 270404 (2001).
\bibitem{olevoasymptotic} A. S. Holevo, quant-ph/0307225 (2003).
\bibitem{refframe} G. Chiribella, G. M. D'Ariano, P. Perinotti, and M. F. Sacchi,
  Phys. Rev. Lett. {\bf 93} 180503 (2004)  
\bibitem{phase} G. M. D'Ariano, C. Macchiavello, P. Perinotti, Phys. Rev. A (in press)
\bibitem{covmeas} G. Chiribella, G. M. D'Ariano, P. Perinotti, and M. F. Sacchi, Phys. Rev A {\bf  70} 062105 (2004) 
\bibitem{infolocvsglob} G. M. D'Ariano, P. Perinotti, and M. F. Sacchi,  Phys. Rev. A {\bf 72} 042108 (2005)
\bibitem{Ballester} E. Bagan, M. A. Ballester, R. Mu\~noz-Tapia, O. Romero-Isart,
  Phys. Rev. Lett. {\bf 95} 110504 (2005) 
\bibitem{Ballester2} M. A. Ballester, quant-ph/0506197 (2005)
\bibitem{Ballester3} E. Bagan, M. A. Ballester, R. Munoz-Tapia, O. Romero-Isart quant-ph/0505083
\bibitem{Ballester4} M. A. Ballester, Phys. Rev. A 70, 032310 (2004)
\bibitem{Vittorio} V. Giovannetti, S. Lloyd, and L. Maccone, quant-ph/0509179 (2005)
\end{thebibliography}
\end{document}